\documentclass[dvips]{article}
\usepackage{icrctc07}

\title{The Lunar Cherenkov Technique: From Parkes Onwards}
\shorttitle{The Lunar Cherenkov Technique}

\authors{C. W. James$^{1}$,  R. D. Ekers$^{2}$, R.~A.~McFadden$^3$, R.~J.~Protheroe$^{1}$}
\shortauthors{C.~W.~James et al.}
\afiliations{$^1$ School of Chemistry \& Physics, Univ.\ of Adelaide, SA, Australia \\ $^{2}$Australia Telescope National Facility, Epping, NSW, Australia\\
$^3$ University of Melbourne, Vic., Australia}
\email{clancy.james@adelaide.edu.au}

\abstract{The lunar Cherenkov technique, which aims to detect the coherent Cherenkov radiation produced when UHE particles interact in the lunar regolith, was first attempted with the Parkes radio-telescope in 1995, though the theory
was not sufficiently developed at this time to calculate a limit on the UHE neutrino flux from the non-observation.
Since then, the technique has evolved to include experiments utilising lower frequencies, wider bandwidths, and entire arrays of antenna.
We develop a simulation to analyse the full range of experiments, and calculate the UHE neutrino flux limit from the Parkes experiment, including the directional dependence. Our results suggest a methodology for planning future observations, and demonstrate how to utilise all available information on the nature of radio pulses from the Moon for the detection of UHE particles.}

\begin{document}
\maketitle
\section{Introduction}

The lunar Cherenkov technique aims to detect ultra-high energy (UHE: $>$EeV) cosmic rays and neutrinos via the coherent Cherenkov radiation emitted in the radio regime by their interactions in the outer layers of the Moon. The radiation is detected via ground-based radio-telescopes, as first proposed by Dagkesamanskii \& Zheleznykh \cite{lunatic_original} and attempted at Parkes \cite{parkes_original}.

The use of existing radio-telescopes has meant that the choice of observational parameters such as frequency, bandwidth, and length of run has been limited. A `next-generation' radio-array, the SKA (Square Kilometre Array, \cite{spie_SKA}), is currently in the planning stages, providing a unique opportunity to optimise a powerful instrument for this technqiue. The ultimate goal of the lunar Cherenkov technique is to determine the energy and arrival directions of the highest energy cosmic rays and neutrinos. Here, we propose how observations might be tailored to achieve this, by examining the frequency- and directional-dependence of the sensitivity of both the Parkes experiment and a nominal SKA to UHE neutrinos.



\subsection{Simulations}
\label{sims}

The program used to simulate the lunar Cherenkov technique was a Monte Carlo code described in James et al.\ \cite{parkes_limits}. This code, like previous simulations, only included a single layer in which detectable UHE particle interactions were allowed. However, as the regolith properties are expected to reflect that of the underlying material, we modified the code to allow UHE neutrino interactions in the sub-regolith layer, modelled with density $3$~g/cm$^3$. This layer was treated simply as denser regolith, with refractive index $n$ and absorption length $\ell$ scaled by density as per Olhoeft \& Strangway \cite{dielectric_properties} to $n=2.5$ and $\ell=29~\lambda$.

The effect of including only the classical regolith was to artificially limit the aperture to neutrinos of energies sufficiently large that interactions at the base of this (typically $10$~m deep) layer were being readily detected, especially at low frequencies, where attenuation within the regolith is minimal. Including two layers is more appropriate than extending a single layer to greater depths, since some radiation losses are to be expected during the transmission of radio-waves from the underlying material (be it lava flows in the mare or megaregolith in the highlands) to the regolith.

The parameters used to simulate the Parkes experiment are also described James et al. For the nominal SKA, we use an effective collecting area of $A_{\rm eff}=1$~km$^2$, a system temperature $T_{\rm sys}$ = $50^{\circ}$~K, and uniform coverage of the Moon over all frequencies. Detection required an integrated signal strength in V/m over the full bandwidth of $8$ times the RMS noise voltage.

\subsection{Results -- Frequency Dependence}
The first lunar Cherenkov experiments observed at frequencies in the $>1$~GHz range, as the power radiated at the Cherenkov angle scales with $\nu^2$ until the turnover frequency ($2.3$~GHz in the regolith \cite{amz_latest}), above which decoherence effects from within the particle cascade become important. Recently, focus has been on observations at lower frequencies, as proposed by Scholten et al.\ \cite{scholten_optimal}. At low frequencies, the Cherenkov cone is broader, and the radiation suffers less attenuation in propagating to the surface, allowing signals from a wider range of interaction geometries at greater depths in the regolith to be detected. For a simulated experiment with LOFAR, assuming a bandwidth of $20$~MHz, Scholten et al.\ found an optimum frequency range in the $115-240$~MHz band, below which (in the $30-80$~MHz band) the effect of increasing sky temperature on sensitivity became significant.

High-frequency experiments have a lower energy detection threshold, since the signal from geometrically favourable events peaks in the GHz regime. How this impacts upon event rates depends on the shape of the UHE particle spectrum, and the aperture at these energies. Another reason to not limit observations to low frequencies only is that to perform true UHE particle astronomy, the type, energy, and arrival direction of any detected particles will need to be determined. The sensitivity of the high frequency component to interaction geometry means it could be used to resolve ambiguities between interaction energy, interaction depth, and orientation of the shower axis with respect to the observer, which would likely remain unresolved in an experiment with a low frequency range.

\begin{figure}
\begin{center}
\noindent
\includegraphics [width=0.33\textwidth, angle=270]{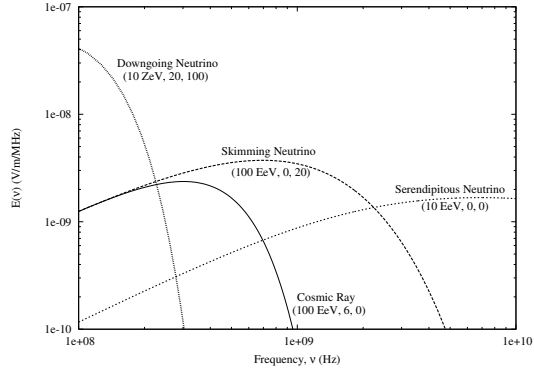}
\vskip -2mm
\end{center}
\caption{\small Simulated spectra $E(\nu)$ of four different particle interactions as viewed from Earth. The three parameters $(E,d,\theta)$ give interaction energy, depth (m), and viewing angle from the Cherenkov angle respectively.}
\label{spectra}
\end{figure}

The range of spectra plotted in Fig.\ \ref{spectra} shows the dilemma. Any single choice of bandwidth over which to trigger results in reduced sensitivity to some fraction of events, since noise from the frequency range over which the signal is weakest will force a higher detection threshold. To detect the full range of events, it will be essential to observe across the entire frequency range. However, the design of an appropriate trigger algorithm is non-trivial, and as discussed by McFadden et al.\ \cite{Rebecca}, is a work in progress. Here we restrict our results to `small' ($\Delta \nu = 0.3 \, \nu$) bandwidths, which necessarily underestimates the SKA's potential, but gives a good indication of the relative importance of the frequency ranges examined.

\begin{figure}
\begin{center}
\noindent
\includegraphics [width=0.33\textwidth, angle=270]{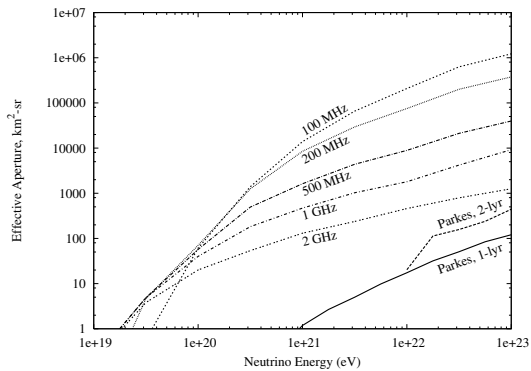}
\vskip -4mm
\end{center}
\caption{\small Calculated aperture to UHE neutrinos of a nominal SKA observing at various frequencies. Also shown is the Parkes aperture both with (this paper) and without (\cite{parkes_limits}) sub-regolith layers included.}
\label{apps}
\end{figure}

We run our simulation for the range of frequencies from $100$~MHz to $2$~GHz, using bandwidths set at a fixed fraction ($30\%$) of the central frequency, reflecting that instruments designed to observed at higher frequencies tend to have a higher available bandwidth. The results, together with the recalculated Parkes aperture, are shown in Fig.\ \ref{apps}.

For neutrino energies $>10^{21}$~eV (i.e.\ well above threshold), and frequencies between $140$~MHz and $1$~GHz, the total effective aperture varies approximately as $\nu^{-3}$, in agreement with that of Scholten et al. However, this dependence weakens at lower energies, and as neutrino energy decreases below $10 \times$ the effective detection threshold, the aperture peaks at higher and higher frequencies. Therefore, the choice of an optimal aperture depends on the assumed UHE neutrino flux, about which little is known.


\subsection{Results -- Directional Sensitivity}
\label{results_directional}

\begin{figure}[b!ht]
\begin{center} \noindent
\vskip -3mm
{\hskip -3mm \includegraphics [width=0.48\textwidth]{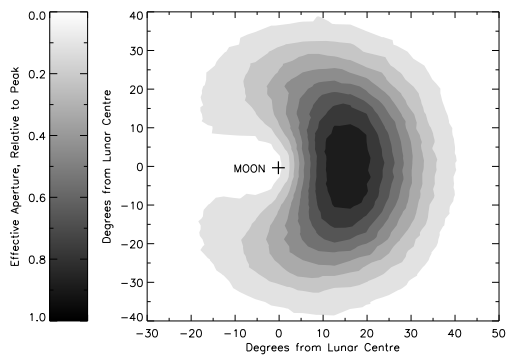}}
\vskip -5mm
\end{center}
\caption{ \small Directional sensitivity of the Parkes experiment to $10^{22}$~eV neutrinos in limb-pointing mode. The Moon is at $(0,0)$, beam centre at $(0.5^{\circ},0)$.}
\label{instantapp} \end{figure}

Including geometrical tags in the simulation allows the sensitivity to be determined as a function of particle arrival direction, relative to the lunar center and beam pointing position. Fig.\ \ref{instantapp} shows the instantaneous aperture of the Parkes experiment in limb-pointing configuration. The total fraction of the sky seen at any one instant is small; the Moon's opacity to neutrinos at UHE creates a `shadow' much larger than its apparent size ($0.5^{\circ}$) on the (left) side opposite the beam pointing position.

We define the `directionality', $d$, to be the ratio of peak over mean directional sensitivity, which measures the ratio of potential point-source sensitivity over the sensitivity to an isotropic flux.
For Parkes in limb-pointing configuration, this is 28; should a point-like neutrino-source be suspected/discovered, a careful choice of observation times and beam pointing positions would enable a much stronger limit to be placed.

\begin{figure}[thb]
\begin{center}
\noindent
\vskip -3mm
\scalebox{0.4}{ \includegraphics*[2.5cm,11.5cm][20cm,24cm]{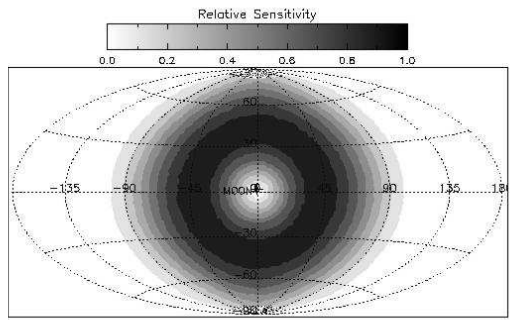}}
\vskip -8mm
\end{center}
\caption{\small Normalised instantaneous aperture (relative units) of a nominal SKA (see text) observing from $100-150$~MHz to $10^{21}$~eV neutrinos. The plot is over the entire sky, with the Moon at the centre.}
\label{lowfinstant}
\end{figure}

For experiments (eg.\ with the SKA) which observe the Moon uniformly, the sensitivity forms an annulus around the Moon, as shown in Fig.\ \ref{lowfinstant}. Similar results would be obtained by placing multiple smaller beams around the limb of the Moon, as might be achieved with a focal plane array. Unlike the total aperture, the directional properties vary slowly with observation frequency, with the annular sensitivity increasing in width with reduced frequency. An important result is that for energies well above threshold, the aperture to all arrival directions is greater at low frequencies.

\begin{figure}
\begin{center}
\noindent
\scalebox{0.38}{ \includegraphics*[2.5cm,13.8cm][20cm,24.5cm]{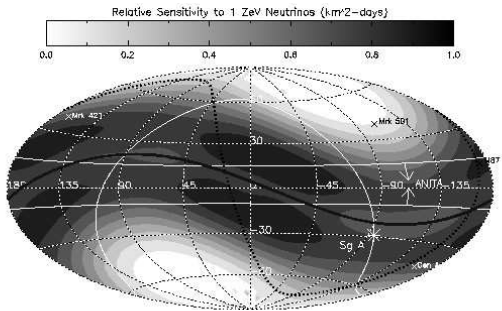}}
\vskip -5mm
\end{center}
\caption{\small 
Relative directional sensitivity of a $350-490$~MHz SKA located at the equator to $1$~ZeV neutrinos (contour increment $0.1$). Observations were spread uniformly over a lunar month. Also shown is the peak ANITA sensitivity range ($\sim 99\%$ within $-10^{\circ} < \delta < +15^{\circ}$ \cite{anita_dec}), the galactic (solid white line) and supergalactic (dotten black line) planes, the apparent lunar orbit (thick black line), and the positions of some notable AGN, used as proxies for point-like UHE neutrino sources.}
\label{skymap}
\end{figure}

Fig.\ \ref{skymap} plots the long-term directional sensitivity to $10^{21}$~eV neutrinos of a low- to mid-frequency SKA ($350-490$~MHz) placed at the equator.
Evidently, there exists a significant overlap between the nominal ANITA and SKA apertures, and large regions of the sky to which neither experiment is sensitive. SKA coverage will be more uniform at lower frequencies. Fortuitously, the sinusoidal `wobble' of the lunar orbit about $\delta=0^{\circ}$ has phase aligned approximately with the supergalactic plane, from which we might expect to see an excess of UHE particles. Since the SKA will be sited in either South Africa or Australia, the sensitivity will be greater to any Southern Hemisphere sources (eg the Galactic Centre), since the Moon will be visible for longer when it is in negative declinations.

\subsection{Discussion}

The highly directional sensitivity of the Parkes experiments suggests that current limits on the UHE neutrino flux are a strong function of celestial coordinates. The current strongest limits on an isotropic flux below $10^{23}$~eV come from ANITA-lite \cite{anita-lite}, which had little sensitivity beyond $15^{\circ}$ from zero declination. Limits from other lunar Cherenkov experiments observing at higher frequencies at Goldstone \cite{goldstone} and Kalyazin \cite{kalyazin} will be more directional than those at Parkes, and thus large areas of the sky would still be left unprobed. This leaves ample scope for experiments with different directional sensitivities to make useful observations to further constrain the UHE neutrino flux.

The future of the lunar Cherenkov technique as a means to determine the flux of the highest energy cosmic rays and neutrinos hinges on the ability to turn positive detections into useful information on particle energy and arrival direction. This requires taking full advantage of the wide frequency range available to instruments such as the SKA, and will likely require multiple detection algorithms, optimised for various pulse spectra, to be used in parallel. The optimisation must take into account the steepness and degree of anisotropy in the (known or expected) UHE particle flux, as well as the sensitivites of other experiments. The already high signal processing requirements of de-dispersion and sub-nanosecond time resolution over a large array will make further innovation in pulse detection technology a necessity.

\subsection{Conclusion}
We have found the lunar Cherenkov technique to be highly directionally sensitive, especially at high frequencies. We have also demonstrated that observations over a broad frequency range with the SKA will be a very powerful technique for UHE particle astronomy. With further improvements in simulation methods, and knowledge of the lunar near-surface, it will be possible to devise advanced pulse detection and reconstruction algorithms, based on the methodology presented here.

\subsection{Acknowledgements}
This research was supported under the Australian Research Council's Discovery funding scheme (project \# DP0559991). Professor R. D. Ekers is the recipient of an Australian Research Council Federation Fellowship (project \# FF0345330).

\end{document}